
\magnification=1200
\vfill
\line{\hfil UB-ECM-PF-09}
\line{\hfil May 1995}
\vskip 2truecm
\font\gran=cmbx12
{\centerline{\gran Combining MCRG and Fourier Accelerated Langevin
Algorithm}}

\vskip 1truecm
\bigskip
\centerline{D. Espriu\footnote{$^*$}{e-mail: espriu@greta.ecm.ub.es}  ~and~
A. Travesset\footnote{$^\dagger$}{e-mail: alex@greta.ecm.ub.es}}
\centerline{\it D.E.C.M., Facultat
de F\'\i sica and I.F.A.E.} \centerline{\it Universitat de Barcelona}
\centerline{\it Diagonal, 647}
\centerline{\it E-08028 Barcelona}
\vskip 1.5truecm
\bigskip
\bigskip
\centerline{ABSTRACT}
\bigskip
We study the implementation of Monte Carlo renormalization group (MCRG) in
momentum
space. This technique is most efficient when used  in combination
with a Fourier accelerated Langevin algorithm. As a benchmark we calculate
the critical exponents $\nu$ and $\eta$ in the vicinity of both
the gaussian and the Wilson
fixed point in $\lambda \phi^4_3$. The results are very competitive with
alternative analytical methods and require a moderate computational
effort only.
\vfill
\eject

\beginsection{\bf 1. Introduction}

 Montecarlo renormalization group[1] is a powerful
technique to determine
critical exponents in numerical simulations. In
this paper we implement this technique in momentum space
(as opposed to real space)
in conjunction
with a Fourier accelerated Langevin algorithm, used to
generate the set of field configurations upon which the
renormalization group transformation is performed. Our
primary interest is
its application to the crumpling
transition in random surface models[2] but we have found the
results for a $Z_2$-symmetric,
one-component, scalar field theory in 3 dimensions
interesting in their own right since most MCRG studies have been
performed with spin models. These results are reported here.

Fourier
accelerated algorithms were introduced by Batrouni et al. [3].
The method uses a stochastic Langevin equation in momentum space, partially
eliminating critical slowing down. Being already in momentum space it is
natural to define a Kadanoff transformation there,
discarding the high frequencies of the system
and keeping
the low frequencies, which are the relevant ones near
criticality. The procedure is close in spirit to Wilson's original
proposal[4] and it apparently avoids some difficulties that arise with
some `unphysical' renormalization group transformations in real space.
The method we discuss also possesses the desirable property of being
exact in the free case. And, last but not least,
the present method delivers very accurate results for the critical
exponents $\nu$ and $\eta$ at the Wilson fixed point with little
computational effort.
Subleading exponents at the Wilson
fixed point
are of course harder to get and
we do not have accurate values for them yet, but they
do not seem to be hopelessly beyond reach.
\bigskip

\line{\bf 2. Numerical Renormalization Group.\hfil}
\medskip\noindent
Just to fix the notation, let us briefly review the
method. Our starting point is the bare action
$$ S =\sum_{\alpha} \lambda_{\alpha}^b O_{\alpha}^b\eqno(2.1) $$
If we perform a renormalization group transformation with a scale
$t$ we end up with a similar action, but with renormalized
couplings and operators.
The
new couplings will be highly nonlinear functions
of the old ones
$$ \lambda_{\alpha}^r=F(\lambda_{\alpha}^b)\eqno(2.2) $$
Near the fixed point we can linearize such function
$$  \lambda_{\alpha}^r-\lambda_{\alpha}^{\ast} =\sum_{\beta}
    T_{\alpha \beta}\left(\lambda_{\beta}^b
-\lambda_{\beta}^{\ast} \right) \qquad
T_{\alpha\beta}=\left({\partial\lambda_
\alpha^r}\over{\partial\lambda_\beta^b}\right)_
{\lambda^b=\lambda^\ast}\eqno(2.3)$$
The largest eigenvalue $\lambda_h$ of $T_{\alpha \beta}$ will
give the critical exponent $\nu$
$$ \nu = \ln t/\ln \lambda_h\eqno(2.4) $$
To compute $T_{\alpha
\beta}$ we use the chain rule
$${{\partial\langle O_{\gamma}^r\rangle}\over{\partial
\lambda_\beta^b}}=
  \sum {\left({\partial\lambda_
\alpha^r}\over{\partial\lambda_\beta^b}\right){
{\partial\langle O_{\gamma}^r\rangle}
 \over{\partial \lambda_\alpha^r}}}\eqno(2.5) $$
The expectation value of a renormalized
operator can be computed either directly with the
renormalized action or, using the
renormalization group transformation to express
the renormalized operator in terms of the bare fields,
with the bare action. Thus
$${{\partial\langle O_{\gamma}^r\rangle}
\over{\partial \lambda_\beta^b}}=
\langle O_{\gamma}^r O_{\beta}^b\rangle-
\langle O_{\gamma}^r\rangle \langle O_{\beta}^b\rangle
\eqno(2.6) $$
$${{\partial\langle O_{\gamma}^r\rangle}
\over{\partial \lambda_\beta^r}}=
\langle O_{\gamma}^r O_{\beta}^r\rangle -\langle O_{\gamma}^r\rangle
\langle O_{\beta}^r\rangle
\eqno(2.7) $$
All one needs is a set of statistically independent field
configurations to determine these correlators and, in turn,
$T_{\alpha\beta}$.
MCRG has been most successful when applied
to Ising spins in 2 and 3 dimensions[1,5]. See [6] for
other applications.

The
Kadanoff transformation is largely arbitrary. We shall consider
a transformation with $t=2$, reducing the volume of our system
from $ N=n_1\times n_2\times ...\times n_d $ points to
$ N_s=n_1/2\times n_2/2\times...\times n_d/2 $. The renormalization
group transformation we will use to achieve this volume reduction will be a
decimation in momentum space. At each step we will simply discard the high
$p$ modes.
This transformation is very easy to implement
numerically and it has many advantages, as it will be clear in
what follows.

A renormalization group transformation is not just a Kadanoff
transformation. It also involves a
field rescaling because redefinitions of the
fields entering in the action make the function (2.2) ambiguous.
We must pick an operator (usually the kinetic term) to have
a definite normalization which will be preserved.
Let us work out this rescaling. For a free theory the action is
$$ S={1\over 2}\sum_{x,\mu}(\phi(x)-\phi(x+\hat{\mu}))^2 +
{m^2\over 2 } \sum_x\phi(x)^2 \eqno(2.8)$$
We set the lattice spacing equal to 1 and assume
a hypercubic lattice in $d$ dimensions.
Define
$$ \phi(x)={1\over N}\sum_p \exp(-i \vec p \vec x)
\phi(p)\eqno(2.9) $$
then
$$ S={1\over{2N}}\sum_{p} \phi(- p) \left( 2d -
2\sum_{\mu=1}^d\cos p_\mu +m^2 \right)\phi( p)\eqno(2.10) $$
We now discard the modes with $ \vert p_\mu \vert > \pi/2 $
and re-express the action in real space via (2.9), but on the small
lattice of size
$N_s=N/2^d $. The result is [4]
$$ S= {1\over 2} {N_s\over{4N}}\sum_{x,\mu}(\phi(x)-
 \phi(x+\hat{\mu}))^2 +{m^2\over 2}{N_s\over N}  \sum_x\phi(x)^2
\eqno(2.11)$$
To get the same kinetic term we need to redefine
$\phi(x)=\zeta \phi(x)$ where
 $\zeta=1/2^{(d+2)/2}$. Then the only other renormalized coupling ---the
mass--- is $m^2_r=4m^2$, as expected. In an interacting theory we
cannot determine the rescaling $\zeta$ exactly. We shall write
$\zeta = 1/2^{(d+2-\eta)/2}$
where $\eta$ is the deviation in the
scaling of the kinetic term with respect to the free case. Provided
we are close enough
to a fixed point, $\eta$ is the anomalous dimension of the field
$\phi$.

We shall determine $\zeta$ (and, consequently, $\eta$)
selfconsistently by demanding  that observables computed with the
bare and renormalized actions should coincide. The results are
quite sensitive to the value of $\eta$. Departures from the correct
value will quickly lead to unacceptable values for the eigenvalues
of the $T$ matrix (very large, complex, ...).

\bigskip
\line{\bf 3. The Fourier Accelerated Langevin Algorithm (FALA).\hfil}
\medskip
One alternative to (partially) beat critical slowing down in Monte Carlo
methods is to
use the Langevin equation to construct a random walk in configuration space
$$ \phi(x,t_{n+1})=\phi(x,t_{n})+\Delta \phi(x,t_n)\eqno(3.1) $$
$$ \Delta \phi(x)=-\Delta t \sum_y \epsilon(x,y) {{\delta S}\over{\delta
\phi(y,t_n)}}+ \sqrt{\Delta t} \eta(x,t_n)\eqno(3.2) $$
where $\eta(x,t)$ is a gaussian noise satisfying
$$ \langle\eta(x,t) \eta(y,t')\rangle = 2 \delta_{t,t'}
\epsilon(x,y)\eqno(3.3) $$
In momentum space the Langevin equation reads
$$ \phi(p,t_{n+1})=\phi(p,t_{n})-\Delta t \epsilon(p)
{\cal F} {{\delta S}\over{\delta \phi(y,t_n)}}+\sqrt{{\Delta t}
\epsilon(p)} \eta(p,t_n)\eqno(3.4) $$
and
$$ \langle\eta(p_1,t) \eta(p_2,t')\rangle = 2 \delta_{t,t'}
\delta{p_1,-p_2}\eqno(3.5) $$
Given infinite computer resources $\epsilon(p)$ is irrelevant.
Any will do. However, to beat critical slowing down a physically
motivated choice is called for. For the free theory, using
$\epsilon(p)=D(p,m)$ ($D$ is the free lattice propagator) is the
optimal choice ---all modes are updated at the same rate[3].
We expect that in the interacting case a good choice will be
$\epsilon(p)=D(p,1/\xi)$, where
$\xi$ is the correlation length in units of
the lattice spacing.

One point to worry about is that for finite $\Delta t$ the
equilibrium distribution is not the one corresponding to
$S$ but rather to
$$ S_{eq}= S +{1\over 4}\sum_{xy}\Delta t\epsilon(x,y)
({{\delta^2 S}\over {\delta \phi(x) \delta \phi(y)}}
-{{\delta S}\over {\delta \phi(x)}}{{\delta S}\over {\delta \phi(y)}})
+{\cal O}(\Delta t)^2\eqno(3.6)$$
Our action $S$ will just be (2.8) with the additional term
$${\lambda\over 4}\sum_x \phi(x)^4\eqno(3.7)$$
The finiteness of the Langevin step
induces a systematic error of ${\cal O}( \Delta t)$ in $S_{eq}$.
For instance, the bare mass gets renormalized by an amount
that is approximately linear in $\Delta t$.
We would therefore be inclined to use a value for $\Delta t$ as
small as possible, but this requires longer runs to reduce the
statistical error.

The latter  can be estimated as
$\sqrt{2\tau / N_i}$, $N_i$ being the total number of Langevin
iterations, fixed by budgetary constraints, and $\tau$
being the autocorrelation time, fixed by $\Delta t$, but depending on
the model dynamics too. For the free theory $\tau\sim 1/\Delta t$ (the
actual
proportionality constant depends on the normalization of $\epsilon(p)$).
Away from
the critical point this relation will hold for the interacting theory too;
the autocorrelation time will be proportional to $(\Delta t)^{-1}$ and
approximately $\lambda$ independent.
However, as we approach criticality the proportionality constant
in the autocorrelation
time will grow as $\xi^z$ ($\xi $ being the correlation length). For
local Monte Carlo
algorithms $z=2$, while FALA are
`optimal' having $z=0$ for multicomponent scalar theories. In our case,
one-component scalar $\lambda\phi^4$, the presence of a discrete
move $\phi\to -\phi$ degrades the performance to $z\simeq 1-2$[7]. The
behaviour of the autocorrelation time is shown in Fig. 1. The difficulty
with the discrete move will be absent in most applications.

Adding the two errors in quadrature we get that the optimal
Langevin time step is proportional to $N_i^{-1/3}$.
Trial and error also leads to Langevin steps
in rough agreement with this scaling.
A possibility to decrease the systematic error is to use a
second order Langevin algorithm effectively removing the
${\cal O}(\Delta t)$ corrections from $S_{eq}$. The resulting code
is of course slower, poorer statistics partially making up for any gains
in precision. Our conclusion after some preliminary runs is that it
hardly pays to use a second order algorithm.

\bigskip
\line{\bf 4. The Gaussian Fixed Point.\hfil}
\medskip\noindent
This corresponds to the point $\lambda=0$, $m^2=0$. Since we set
$\lambda=0$ from the start, the results we will present are trivial to
obtain analytically, but they are a test that any MCRG method should
pass. Autocorrelations are reduced completely
(i.e. $\tau\simeq 1$)
if $\Delta t \simeq 5\times 10^{-2}$ (even though such big Langevin step
induces a renormalization on the fields and the mass itself).
We have checked that
Wick's theorem is satisfied with very good accuracy.
10 bins with $5\times 10^3$
configurations per bin were enough to get accurate results.
We have used a lattice of volume $32^3$ reducing it to
$16^3$. The rule-of-thumb for $\Delta t$ previously discussed
suggests $\Delta t=7\times 10^{-4}$ which is the value we have used.
The operators we have considered
are
$ O_n(\phi)=\sum_x \phi(x)^{2n} $.
and their respective dispersions $\sigma(O_n)$.
No `derivative' operators
were considered at this stage.

A first application of the algorithm taking $m^2=10^{-3}$
gives the results shown in Table 1. We can extract the
critical exponent $\nu=.5027(30)$ (exact $\nu=.5$). The second
eigenvalue is not very accurate. If we are very near a fixed point the
renormalized and bare action are the same so averages of observables
should coincide, that is
$ \langle O^b_n \rangle = \langle O^r_n \rangle$.
In  this way, by fitting some observables, we
can get the anomalous dimension $\eta$. In our case, including an
anomalous
dimension worsens the agreement, so we take $\eta = 0$.

If we gradually approach the critical point by letting $m^2
\to 0$ the observables match and the subleading
critical exponents agree with their exact values (Tables 3 and 4). Our
best result for the critical exponent is
$\nu=.5004(5)$. Going to much smaller masses is problematic
due to finite
size effects.
The results are remarkably stable with respect to the addition or
removal of higher dimensional operators.
The lesson we extract from this simple
example
is that it it is easy to get the first eigenvalue right, but to get the
subleading critical exponents one must be very close to the fixed point.

The errors quoted in the critical exponents are overwhelmingly dominated
by systematic effects due to the truncation of the bare and renormalized
actions. Statistical errors are negligibly small in all cases presented.
Errors due to the fact that $S_{eq}\neq S$ are naively expected to cancel
out in universal quantities. However, regarding this point one must keep
in mind that, because a finite Langevin step modifies in
practice the action of the system, $S_{eq}$ is no longer at the gaussian
fixed point.
Then matching between bare and
renormalized expectation values is no longer guaranteed (of course, since
$\Delta t$ is very small, the matching is excellent for $m^2=10^{-5}$, but
not perfect).
For this reason, including many operators in the renormalized action does
not automatically improve the results; it may actually slightly worsen
them because small errors in the highest dimensional operators eventually
affect the precision of the largest eigenvalues. Finally,
errors due to finite size effects
do not seem to be relevant for these values of $m^2$.

$$\vbox{\tabskip=0pt \offinterlineskip
\halign to280pt{\strut#& \vrule#\tabskip=1em plus2em&
\hfil#& \vrule#& \hfil#\hfil& \vrule#& \hfil#& \vrule#\tabskip=0pt\cr
\noalign{\hrule} & & \multispan5\hfil Table 1: $m^2=10^{-
3}$,  $\lambda =0$ \hfil& \cr
\noalign{\hrule} & & \omit\hidewidth Operators\hidewidth& &
\omit\hidewidth 1 eigen \hidewidth& & \omit\hidewidth
2 eigen \hidewidth& \cr
\noalign{\hrule} & &1&      &3.993&  &------& \cr
\noalign{\hrule} & &2&      &3.977&  &1.688& \cr
\noalign{\hrule} & &3&      &3.972&  &1.775& \cr
\noalign{\hrule} & &4&      &3.964&  &1.731& \cr
\noalign{\hrule} & &5&      &3.967&  &1.731& \cr
\noalign{\hrule} & &6&      &3.967&  &1.731& \cr
\noalign{\hrule} & &7&      &3.964&  &1.784& \cr
\noalign{\hrule} & &8&      &3.967&  &complex& \cr
\noalign{\hrule} & &exact&  &4.000&  &2.000& \cr
\noalign{\hrule}\noalign{\smallskip}}}$$

$$\vbox{\tabskip=0pt \offinterlineskip
\halign to280pt{\strut#& \vrule#\tabskip=1em plus2em&
\hfil#& \vrule#& \hfil#\hfil& \vrule#& \hfil#& \vrule#\tabskip=0pt\cr
\noalign{\hrule} & & \multispan5\hfil Table 2: $m^2=10^{-
3}$,  $\lambda =0$ \hfil& \cr
\noalign{\hrule} & & \omit\hidewidth $O_n/N$ \hidewidth& &
\omit\hidewidth big lat. \hidewidth& & \omit\hidewidth
small lat. \hidewidth& \cr
\noalign{\hrule} & &1&      &0.259&  &0.219& \cr
\noalign{\hrule} & &2&      &0.202&  &0.144& \cr
\noalign{\hrule} & &3&      &0.262&  &0.157& \cr
\noalign{\hrule} & &4&      &0.475&  &0.239& \cr
\noalign{\hrule} & &5&      &1.104&  &0.465& \cr
\noalign{\hrule} & &6&      &3.141&  &1.103& \cr
\noalign{\hrule} & &7&      &10.59&  &3.077& \cr
\noalign{\hrule} & &8&      &41.47 &  &9.800& \cr
\noalign{\hrule}\noalign{\smallskip}}}$$

$$\vbox{\tabskip=0pt \offinterlineskip
\halign to280pt{\strut#& \vrule#\tabskip=1em plus2em&
\hfil#& \vrule#& \hfil#\hfil& \vrule#& \hfil#\hfil& \vrule#&
\hfil#\hfil& \vrule#& \hfil#& \vrule#\tabskip=0pt\cr
\noalign{\hrule} & & \multispan9\hfil Table 3: $m^2=10^{-
5}$,  $\lambda =0$ \hfil& \cr
\noalign{\hrule} & & \omit\hidewidth Op.\hidewidth& &
\omit\hidewidth 1 eig \hidewidth& & \omit\hidewidth
2 eig \hidewidth& & \omit\hidewidth 3 eig \hidewidth & &
\omit\hidewidth 4 eig \hidewidth& \cr
\noalign{\hrule} & &1& &3.999& &------& &------& &------&\cr
\noalign{\hrule} & &2& &3.997& &2.003& &------& &------&\cr
\noalign{\hrule} & &3& &3.994& &2.005& &1.003& &------&\cr
\noalign{\hrule} & &4& &3.994& &1.985& &1.003& &0.402&\cr
\noalign{\hrule} & &5& &3.993& &1.999& &0.999& &0.551&\cr
\noalign{\hrule} & &6& &3.994& &1.989& &1.019& &0.553&\cr
\noalign{\hrule} & &7& &3.994& &1.998& &0.964& &0.584&\cr
\noalign{\hrule} & &8& &3.994& &2.002& &0.930& &0.620&\cr
\noalign{\hrule} & &exact& &4.000& &2.000& &1.000& &0.500&\cr
\noalign{\hrule}\noalign{\smallskip}}}$$

$$\vbox{\tabskip=0pt \offinterlineskip
\halign to280pt{\strut#& \vrule#\tabskip=1em plus2em&
\hfil#& \vrule#& \hfil#\hfil& \vrule#& \hfil#& \vrule#\tabskip=0pt\cr
\noalign{\hrule} & & \multispan5\hfil Table 4: $m^2=10^{-
5}$,  $\lambda =0$ \hfil& \cr
\noalign{\hrule} & & \omit\hidewidth $O_n/N$ \hidewidth& &
\omit\hidewidth big lat. \hidewidth& & \omit\hidewidth
small lat. \hidewidth& \cr
\noalign{\hrule} & &1&      &0.304&  &0.308& \cr
\noalign{\hrule} & &2&      &0.275&  &0.274& \cr
\noalign{\hrule} & &3&      &0.410&  &0.400& \cr
\noalign{\hrule} & &4&      &0.850&  &0.775& \cr
\noalign{\hrule} & &5&      &2.242&  &1.935& \cr
\noalign{\hrule} & &6&      &7.150&  &5.800& \cr
\noalign{\hrule} & &7&      &26.65&  &20.70& \cr
\noalign{\hrule} & &8&      &113.8&  &88.27& \cr
\noalign{\hrule}\noalign{\smallskip}}}$$

\bigskip
\line{\bf 5. The Wilson Fixed Point\hfil}
\medskip\noindent
The critical exponents in the vicinity of this fixed point
are not known exactly. There are many different approaches giving
reasonably precise estimates of the exponents, $\epsilon$ expansion
[8], High
$T_c$ [9],exact Renormalization group treatments [10] and MCRG analysis
[5,11]. The evidence accumulated suggest that no other fixed point
exists and give an average estimate for the exponents

  $$\nu=0.627-0.631 \qquad \eta=0.03-0.04  \eqno(5.1)$$

As we have discussed in section 3 even using a FALA does not
completely suppress critical slowing down all the way to
$z=0$ in the particular example we are considering. To guarantee
sound statistics
we
accumulated $O(10^5)$ configurations in bins of $10^4$, until
observables reach a plateau.
We also measure autocorrelation times. Since the proportionality
constant
relating $\tau$ and $\Delta t$ can be large near the fixed point, the
Langevin step has to be relatively large. The compromise between
statistical and systematic errors leads to $\Delta t\simeq 10^{-3}$.
The observables  get some
small renormalization
but universal quantitities should be relatively unaffected.

We must fine tune the parameters $m^2<0$ and $\lambda$
so that we are near the fixed point. For each $\lambda >0$ there
is a value $m^2_{\lambda}<0$ corresponding to the critical surface.
Applying renormalization group transformations there
would drive us to the fixed point.
Unfortunately we cannot do that in the computer since near the critical
surface the correlation length starts to diverge and finite size effects
enter into the game. What we did is to perform  numerical simulations
for a fixed value of $\lambda$ and then sweep over $m^2$ to find the
critical surface. For this purpose
we simulated sistems with $16^3$, $16^2\times 8$, $16\times 8^2$ and $8^3$
sites. (We were forced
to study asymmetric lattices because fast Fourier transforms require
the
lattice sizes to be a power of 2.) In addition to measuring observables
such as
energy density, specific heat, magnetization, etc. we
also determined the correlation length by fitting
the long distance part of our measured Green functions to a propagator
of the type
$1/{\Delta+\xi^{-2}}$. Once the critical surface is located,
a MCRG is performed at a value of $m^2$ in its vicinity, but where
the correlation length is not too large for our lattice.
MCRG is done on a $32^3$ lattice so the
renormalized lattice becomes $16^3$.
We repeat this procedure for several $\lambda$
values until we observe a near perfect match between the bare
and renormalized actions. This will be the approximate location of the
fixed point. Of course the true fixed point lives in an infinite-dimensional
parameter space and we are simply getting as close to it
as possible in the $(\lambda, m^2)$ plane. Because the agreement
between bare and renormalized expectation values at our approximate fixed
point is good we conclude that the
particular renormalization group transformation we are using has a fixed
point not far from the
$(\lambda,m^2)$ plane.

To begin with, we consider only local operators.
We study values of $\lambda$ ranging from
$0.1$ to 20. For $\lambda=0.1$ the first eigenvalue of the $T$ matrix is
3.969,
almost the one corresponding to the gaussian fixed point. For
such range of parameters we are still in the domain of attraction of the
gaussian fixed point. Then
we increase $\lambda$ gradually repeating the same stratregy. The
largest eigenvalue gradually changes. By the time we reach the
region $5<\lambda<8$ observables in the big and small lattices match
with an accuracy similar to the one around the gaussian fixed point.
In fact, the best agreement is obtained around
$\lambda =6.5$ which we take as approximate location of the new fixed point.
The agreement improves if we allow for a small anomalous
dimension. Fitting the 5 lowest dimensional operators gives
$$ \eta=0.03 \pm 0.005 \eqno(5.2) $$
 The agreement between big and small lattices is illustrated
in Table 5 for the case $\lambda=6$ and Table 6 shows the
results for the critical exponent $\nu$.
Our best estimate is
$$ \nu = 0.628(5) \eqno(5.3)$$
The error corresponds to the error in the eigenvalue. Notice that also
in this case the agreement between expectation values deteriorates
for the highest dimensional operators. This is again a reflection that
our bare action (which contains only three operators) is not quite
at the fixed point, just close to it.
Both $\nu$ and $\eta$ are in perfect agreement with the ones obtained in
[5,8-11].
If we keep increasing $\lambda$  we are still  in the domain of
attraction of the wilsonian fixed point, but we are unable to match
observables in the big and small lattices, even by fitting an anomalous
dimension. Fixing  $\eta=0.04$ we get $\nu = 0.630 (5)$ for $\lambda=10.$,
still in agreement with the value obtained previously.
Table 7
shows the big/small lattice comparison for $\lambda=10.$ with $\eta=0$.

$$\vbox{\tabskip=0pt \offinterlineskip
\halign to280pt{\strut#& \vrule#\tabskip=1em plus2em&
\hfil#& \vrule#& \hfil#\hfil& \vrule#& \hfil#& \vrule#\tabskip=0pt\cr
\noalign{\hrule} & & \multispan5\hfil Table 5: $m^2=-3.32$,
$\lambda =6$ \hfil& \cr
\noalign{\hrule} & & \omit\hidewidth $O_n/N$ \hidewidth& &
\omit\hidewidth big lat. \hidewidth& & \omit\hidewidth
small lat. \hidewidth& \cr
\noalign{\hrule} & &1&      &0.272&  &0.267& \cr
\noalign{\hrule} & &2&      &0.161&  &0.160& \cr
\noalign{\hrule} & &3&      &0.130&  &0.133& \cr
\noalign{\hrule} & &4&      &0.130&  &0.140& \cr
\noalign{\hrule} & &5&      &0.151&  &0.173& \cr
\noalign{\hrule} & &6&      &0.187&  &0.240& \cr
\noalign{\hrule} & &7&      &0.260&  &0.380& \cr
\noalign{\hrule} & &8&      &0.402&  &0.650& \cr
\noalign{\hrule}\noalign{\smallskip}}}$$

$$\vbox{\tabskip=0pt \offinterlineskip
\halign to280pt{\strut#& \vrule#\tabskip=1em plus2em&
\hfil#& \vrule#& \hfil#\hfil& \vrule#& \hfil#& \vrule#\tabskip=0pt\cr
\noalign{\hrule} & & \multispan5\hfil Table 6: $\nu$ for different
$\lambda$ \hfil& \cr
\noalign{\hrule} & & \omit\hidewidth $\lambda$\hidewidth& &
\omit\hidewidth $\nu$ \hidewidth& & \omit\hidewidth
error \hidewidth& \cr
\noalign{\hrule} & &5.0&      &0.623&  &0.006& \cr
\noalign{\hrule} & &6.0&      &0.628&  &0.007& \cr
\noalign{\hrule} & &6.5&      &0.628&  &0.006& \cr
\noalign{\hrule} & &7.0&      &0.628&  &0.005& \cr
\noalign{\hrule} & &7.5&      &0.629&  &0.005& \cr
\noalign{\hrule}\noalign{\smallskip}}}$$

$$\vbox{\tabskip=0pt \offinterlineskip
\halign to280pt{\strut#& \vrule#\tabskip=1em plus2em&
\hfil#& \vrule#& \hfil#\hfil& \vrule#& \hfil#& \vrule#\tabskip=0pt\cr
\noalign{\hrule} & & \multispan5\hfil Table 7: $m^2=-4.94$,
$\lambda =10$. $\eta=0.$ \hfil& \cr
\noalign{\hrule} & & \omit\hidewidth $O_n/N$ \hidewidth& &
\omit\hidewidth big lat. \hidewidth& & \omit\hidewidth
small lat. \hidewidth& \cr
\noalign{\hrule} & &1&      &0.238&  &0.185& \cr
\noalign{\hrule} & &2&      &0.120&  &0.087& \cr
\noalign{\hrule} & &3&      &0.080&  &0.061& \cr
\noalign{\hrule} & &4&      &0.067&  &0.055& \cr
\noalign{\hrule} & &5&      &0.063&  &0.059& \cr
\noalign{\hrule} & &6&      &0.066&  &0.073& \cr
\noalign{\hrule} & &7&      &0.076&  &0.101& \cr
\noalign{\hrule} & &8&      &0.093&  &0.153& \cr
\noalign{\hrule} \noalign{\smallskip}}}$$

We also explored the range of $\lambda$ values up to 20., but the
matching gets much worse and we eventually hit convergence
problems of the Langevin algorithm. It is also important to stress that
no evidence for another fixed point was found, and even for the high
values
of $\lambda$ available the results are in the domain
of attraction of the wilsonian fixed point.

The method delivers $\nu$
(and $\eta$) with good precision. Unfortunately, the second
eigenvalue at $\lambda=6.5$ is 0.3, which corresponds to
$\omega=1.7$, far from the accepted result [10], but these
are subleading effects. One should really get much closer
to the fixed point to be sensitive to these.

It is very important to check the stability of our results under the
inclusion of `derivative' operators. Exact renormalization group
treatments did not include them in their first studies[10a], but some
recent
ones do indeed include some[10b]. We can consider their effect by
including operators of the kind

$$ V_{n}=\sum_{x,\mu}(\phi(x)-\phi(x+\hat{\mu}))^2\phi(x)^{2n} \eqno(5.3)$$
We carefully studied the inclusion of these operators up to $n=6$ in the
computation
of the $T$ matrix in the range $5<\lambda<8$, but the first eigenvalue
did not change at all. However
the second eigenvalue changed
significantly to $\lambda=.82$ (that is $\omega=.28$),
now undershooting the generally accepted value which is $\omega=.5-.7$.
Again the situation in the gaussian fixed point repeats itself, it is
relatively easy to get the first eigenvalue right but the second one
needs being very close to the corresponding fixed point.

\bigskip
\line{\bf 6. Final Remarks.\hfil}
\medskip\noindent
We single out three avenues along which
 the method could be improved.

(1) Iteration of the renormalization group: This would require using
a bigger lattice. On, say, a $64^3$ lattice one could perform two
successive renormalization group transformations. This would surely
improve the matching amongst observables further. In principle,
if our initial bare action is very close to the critical surface, iteration
would bring us closer to the true fixed point. In practice, however, we
cannot be too close to the critical surface due to finite size effects and
while we certainly expect an improvement on our results after iteration this
will not
be a dramatic one, probably. In fact, we find remarkable that our approach
yields such accurate results with so little effort.

(2) Determination of the renormalization group flow trajectories:
The method as it stands so far is not accurate enough to determine the
renormalized couplings. One possibility would be to use the method
suggested in [12]. Another one would require a
combination of MCRG techniques with
a Ferrenberg-Swendsen[13] algorithm. The idea is to use the
Ferrenberg-Swendsen method to perform a numerical simulation in the
small lattice at a value of the parameters where already accurate
results are found, storing as many operators as
possible.
Now apply MCRG paying special attention to the observables that have
been stored.
Then we can compute the  observables for
values of the couplings different from the actual one used in the
simulation, and choose the couplings that best fit the
observables. The couplings define the renormalized action.

(3) Improving the accuracy of subleading exponents:
The fixed point needs to be located rather
precisely, which can be done with the help of the renormalization group
trajectories. Bigger lattices should also improve results because then
we can get closer to the critical surface.

Our results were obtained
with very modest computer resources. All calculations have been carried
out on a IBM Power PC 250. Once the approximate location of the fixed
point is found, the MCRG from a $32^3$ to a $16^3$  system, measuring 14
operators in $10^5$ iterations takes around 50 hours and delivers
critical exponents with a $\sim1\%$ precision.
All the evidence shows that the method is
indeed efficient to study the renormalization group in the lattice. Our
estimates for the critical exponents obtained with quite small lattices
are in perfect agreement with the ones computed by other methods.
We think that the algorithm is well suited to
study more difficult systems ($N$-component $\phi^4$ would be interesting).
The
application to the crumpling transition in random surfaces is now under way.

\bigskip
\line{\bf Aknowledgements.\hfil}
\medskip
It is a pleasure to acknowledge discussions with A.Gonzalez-Arroyo,
J.I.Latorre and E.Moreno. We would like also to thank J.Garcia-Ojalvo
for supplying the gaussian random number
generator.
A.T. aknowledges a Fellowship from the Generalitat
of Catalonia. This work has been supported by contracts CHRX-CT93-0343 and
AEN93-0695.

\bigskip
\line{\bf References.\hfil}
\medskip
\item{[1]}{R.H.Swendsen, Phys. Rev. Lett. 42 (1979) 859;
R.H. Swendsen, in "Real space renormalization",T. Burkhard and J.
Van Leewven ed., Springer Verlag.}
\item{[2]}{P.W. Stephenson and J.F. Wheater, Phys. Lett. B 302 (1993) 447;
J.F. Wheater, Oxford Preprint OUTP-95-07-P}
\item{[3]}{G.G. Batrouni et al, Phys. Rev.D 32 (1985) 2736}
\item{[4]}{K. Wilson and J. Kogut, Phys. Rep. 12 (1974) 75}
\item{[5]}{G.S. Pawley, R.H. Swendsen, D. J. Wallace and K.G. Wilson, Phys
Rev. B 29 (1984) 4030;
C.F. Baillie, K.W. Barnish, R. Gupta and G.S. Pawley, Nucl. Phys.
B (proc. supp.) 17 (1990) 323}
\item{[6]}{C. Rebbi and R.H. Swendsen, Phys. Rev. B21 (1980) 4094;
P. Harten and P. Suranyi, Nucl. Phys. B 265 (1986) 615; A. Hasenfratz et
al.,
Phys. Lett. B 140 (1984) 76; K. Bowler et al., Nucl. Phys. B 257 (1985) 155;
R. Gupta et al., Phys. Lett. B 211 (1988) 132; K. Akemi et al. Phys. Rev.
Lett. 71 (1993) 3063;U. Heller, Phys. Rev. Lett.
60 (1988) 2235; A.Gonzalez-Arroyo, M. Okawa and Y. Shimizu, Phys. Rev. Lett.
60 (1988) 487
\item{[7]}{A.D. Sokal, Nucl. Phys. B (proc. supp.) 20 (1991)
55}
\item{[8]}{J. Zinn-Justin and J.C. Le Guillou, Phys. Rev. B21 (1980) 3976;
G. Parisi, J. Stat. Phys 23 (1980) 49}
\item{[9]}{C. Domb, in "Phase transition and critical phenomena", Vol 3, C.
Domb and M. Green, Academic Press.}
\item{[10a]}{A. Hasenfratz and P. Hasenfratz, Nucl. Phys. B270 (1986) 687;
P.E. Haagensen, Yu. Kubyshin, J.I. Latorre and E. Moreno, Phys.
Lett. B 323, 330 (1994);
T. R. Morris, Int. Jour. Mod. Phys. A9 (1994) 2411;
M. Alford,  Phys. Lett. B 336 (1994) 2411}
\item{[10b]}{G.R. Golner, Phys Rev. B 33 (1986) 7863;
T.R. Morris, Phys. Lett. B 329 (1994) 241;
R. Ball, P. Haagensen, J.I. Latorre and E. Moreno Phys. Lett. B
347 (1995) 80}
\item{[11]}{D.J.E. Callaway and R. Petronzio, Nucl. Phys. B 240 [FS12] 577
(1984)}
\item{[12]}{A. Gonzalez-Arroyo and M. Okawa, Phys. Rev. D35 (1987) 672}
\item{[13]}{A. Ferrenberg and R.H. Swendsen, Phys. Rev. Lett. 61, 2635
(1988)}
\bigskip
\line{\bf Figure Caption.-\hfil}
\medskip
\item{Fig 1.-}{Autocorrelation time for the operator $\phi^2$ at
$\lambda=6$. The Langevin step is $\Delta t=0.001$. The system size is
$16^3$.}
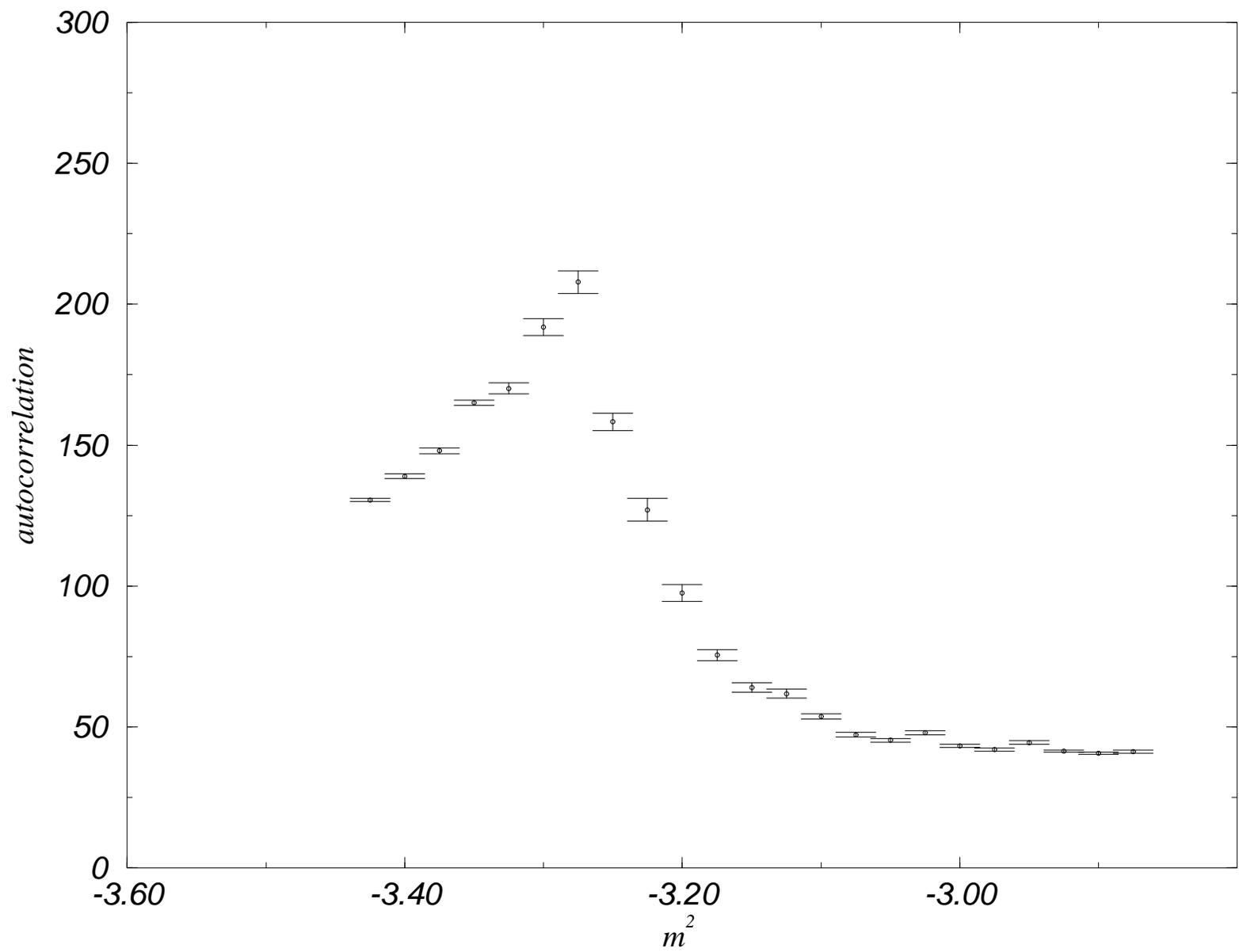
\end